\documentclass[conference, 10pt]{IEEEtran}
\usepackage{cite}
\usepackage{url}

\ifCLASSINFOpdf
\else
 \usepackage[dvips]{graphicx}
  \DeclareGraphicsExtensions{.eps}
\fi

\usepackage{psfrag}

\usepackage[cmex10]{amsmath}

\usepackage{graphicx}
\usepackage{color}
\usepackage{placeins}
\usepackage{float}
\usepackage{tabularx,colortbl}
\usepackage[absolute]{textpos}
\newcommand{\copyrightstatement}{
    \begin{textblock}{15}(0.5,0.3)    
         \noindent
         \centering
         \textblockcolour{white}
         \footnotesize
         \copyright 2016 IEEE. Personal use of this material is permitted. Permission from IEEE must be obtained for all other uses, in any current or future media, including reprinting/republishing this material for advertising or promotional purposes, creating new collective works, for resale or redistribution to servers or lists, or reuse of any copyrighted component of this work in other works
    \end{textblock}
}

\begin{document}
%
\title{Multi-Objective Design Space Exploration for the \\ Optimization of the HEVC Mode Decision Process}
\copyrightstatement

\author{\IEEEauthorblockN{Christian Herglotz, Rafael Rosales, Michael Gla\ss, J\"urgen Teich, Andr\'e Kaup}
\IEEEauthorblockA{Chair of Multimedia Communications and Signal Processing, \\Chair of Hardware/Software Co-Design,\\
Friedrich-Alexander University Erlangen-N\"urnberg (FAU)\\
\{christian.herglotz, rafael.rosales, michael.glass, juergen.teich, andre.kaup\}@fau.de\\}}



\renewcommand{\baselinestretch}{0.9}

\maketitle

\begin{abstract}
Finding the best possible encoding decisions for compressing a video sequence is a highly complex problem. In this work, we propose a multi-objective Design Space Exploration (DSE) method to automatically find HEVC encoder implementations that are optimized for several different criteria. 
The DSE shall optimize the coding mode evaluation order of the mode decision process and jointly explore early skip conditions to minimize the four objectives a) bitrate, b) distortion, c) encoding time, and d) decoding energy.
In this context, we use a SystemC-based actor model of the HM test model encoder for the evaluation of each explored solution.
The evaluation that is based on real measurements shows that our framework can automatically generate encoder solutions that save more than $60\%$ of encoding time or  $3\%$ of decoding energy when accepting bitrate increases of around $3\%$. 
\end{abstract}


\IEEEpeerreviewmaketitle

\section{Introduction}
\label{sec:intro}
During the past two decades, video coding technologies have evolved rapidly.
Each coding standard has introduced new coding tools aiming at reducing the bitrate required to save and transmit the video data.
As a consequence, the latest standard that was finalized in 2013, High-Efficiency Video Coding (HEVC) \cite{Sullivan12} incorporates a tremendous amount of different coding tools that can all be exploited to encode a given sequence as efficiently as possible. 

Due to the high amount of coding tools, the complexity of the encoding process has increased dramatically in comparison to former coding standards.
The HEVC encoder cannot only choose in between different coding modes such as intra- or inter-coding, it also needs to decide for the most suitable block size.
Common solutions like the HM-reference software \cite{HM} usually perform exhaustive searches testing most of the possible coding modes for rate and distortion and choosing the mode with the least rate-distortion cost.
Hence, research aiming at reducing the encoding complexity can help developing efficient and real-time capable encoding solutions.

To this end, in the past couple of years, many publications presented novel and effective encoding methods, such as the mechanisms of Early Skip (ESD)~\cite{jctvc_esd}, Early CU (ECU)~\cite{jctvc_ecu}, and CFB fast method (CFM)~\cite{jctvc_cfm}, which have been included in the HM reference software to speedup the mode decision process. In another work, Zhang et al. \cite{Zhang14b} targeted explicitly the intra mode decision process and investigated how rough mode decisions can help finding a reduced set of mode candidates and achieved a speedup of $2.5$.
In another work, Heindel et al. \cite{Heindel16} proposed analyzing the variance of the intra-reference samples to exclude angular modes when a certain threshold is kept. They achieved encoding complexity savings of up to $50\%$. 
For the inter prediction process, Cao et al. \cite{Cao13} proposed a fast motion estimation scheme for fractional pel interpolation and achieved savings of $33\%$.
Shang et al. \cite{shang2015fast} proposed an analysis approach to perform a fast CTU and PU partitioning through correlations between neighboring CUs and PUs to achieve encoding time savings of $38\%$ on average.
Finally, 
Vanne et al. \cite{Vanne14} investigated how early skipping of modes and the encoding order influence the rate-distortion performance and complexity and proposed a corresponding encoding order.

All the above mentioned work aims at reducing the encoding time or complexity while keeping the bitrate increase at a low level. We adopt this approach as encoding is often performed on desktop PCs, servers, or clusters where fast processing is beneficial. As a second objective, we propose using the decoding energy which is especially interesting for battery-driven portable devices like smartphones or tablet PCs which are nowadays commonly used for video streaming. 
An energy model for estimating the decoding energy has been developed \cite{Herglotz16b} and successfully used to encode decoding-energy saving bit streams \cite{Herglotz16a}, where energy savings of $10\%-20\%$ (at bitrate increases in the same order of magnitude) were reported. 

In our work, we aim at minimizing these two objectives (decoding energy and encoding speed) and the two classic ones, namely bitrate and distortion, jointly. 
As our first main contribution, we propose the optimization of the mode decision process through an automatic multi-objective Design Space Exploration (DSE) of the coding mode evaluation order with respect to these four objectives. 
Second, we evaluate a simple heuristic to explore early skip conditions during the automatic DSE approach.
For the evaluation of different encoding solutions, we take a SystemC-based, actorized model of the HM-16.0 encoder \cite{Rosales16}. 
This actorized HEVC encoder has the advantage of providing the complete set of HEVC prediction modes, as well as to provide the flexibility to customize the evaluation order of the mode decision process, and to specify early skip conditions for each coding mode. Note that although we are using the HM-reference software, the DSE and the resulting solutions are generally applicable to any encoder implementation. 
Finally, in the results section we will show that our approach is capable to automatically find a set of Pareto-optimal encoding solutions nicely characterizing the desired trade-off between bitrate, quality, encoding time, and decoding energy.


The paper is organized as follows: Section \ref{sec:HEVCcoding} introduces the HEVC encoder and its coding modes.
Afterwards, Section \ref{sec:model} introduces actor-based modeling, presents the actorization for the HM software, and shows a proposed genetic representation for design space exploration (DSE) that uses multi-objective evolutionary algorithms~\cite{deb2002fast}. Afterwards, Section \ref{sec:obj} introduces the four objectives and how they are determined in the DSE and the evaluation. 
Then, Section \ref{sec:eval} presents rate-distortion performance, complexity savings, and decoding energy savings in comparison to the HM-reference software using real measurements and interprets the results.
Section \ref{sec:concl} concludes the paper.

\section{HEVC Encoding}
\label{sec:HEVCcoding}
The main goal of a typical encoder is to compress video sequences such that the rate and the distortion is minimized.
To achieve this in HEVC, a frame is divided into so-called Coding Tree Units (CTUs) with a maximum size of $64\times 64$ pixels. Such a CTU can be split recursively into Coding Units (CUs) of size $32\times 32$, $16\times 16$, or $8\times 8$, where the so-called depth $d$ can have a value in between $0$ ($64\times 64$) and 3 ($8\times 8$). Then, for each CU of a given size, the prediction can either be intraframe or interframe.
For the intraframe case, and for a CU size of $8\times 8$, the CU can be further split into four rectangular Prediction Units (PUs). 
For interframe coded CUs, the block can be split into two rectangular PUs, where two symmetric (Nx2N and 2NxN) for all block sizes and four asymmetric partitionings (2NxnD, 2NxnU, nRx2N, nLx2N) for block sizes $64\times 64$, $32\times 32$, and $16\times 16$ are available.

Moreover, each PU can be coded in different coding modes. For our work, the interframe coding modes skip, merge, and inter are of special interest, where in skip mode neither residual coefficients nor a motion vector difference is coded. In merge mode, only the motion vector difference is forced to zero and for the inter mode, none of these constraints hold.


After prediction, the residual error is transformed and quantized.
Furthermore, in-loop filterings (deblocking filter (DBF) and sample adaptive offset (SAO)) can be performed.
An overview on the complete codec and more details on the coding modes can be found in \cite{Sullivan12}.
For this work, especially the CTU partitioning, the interframe prediction modes, and the prediction partitionings are of interest.

The HM encoder exhaustively tests most of the possible modes and chooses the one with the least rate-distortion costs.
Explicit functions are defined calculating this cost for $11$ modes as shown in Table \ref{tab:modes_HM}, where each of the modes introduced above, including a split mode (which means that the current CU is split into four subCUs), is comprised.
\begin{table}[t]
\centering
\caption{Mode evaluation functions in the HM reference software. The right column indicates the depths on which the functions can be executed.}
\label{tab:modes_HM}
\vspace{-0.3cm}
\small
\begin{tabular}{c|l|l l l}
\hline
\# & HM Mode& Tested & Tested & Depth\\
& Function & Coding Modes & PU Partitions\\
 \hline
0 & Intra2Nx2N  & Intra & 2Nx2N & 0,1,2,3\\
1 & Inter2Nx2N  & Inter & 2Nx2N & 0,1,2,3\\
2 & Merge2Nx2N  & Merge, Skip & 2Nx2N & 0,1,2,3\\
3 & InterNx2N  & Merge, Inter & Nx2N & 0,1,2,3\\
4 & Inter2NxN  & Merge, Inter & 2NxN & 0,1,2,3\\
5 & Inter2NxnU  & Merge, Inter & 2NxnU & 0,1,2\\
6 & Inter2NxnD  & Merge, Inter & 2NxnD & 0,1,2\\
7 & InternLx2N  & Merge, Inter & nLx2N & 0,1,2\\
8 & InternRx2N  & Merge, Inter & nRx2N & 0,1,2\\
9 & IntraNxN  & Intra & NxN & 3\\
10 & Split  & All available& All available& 0,1,2\\
& & at lower depth & at lower depth& \\
 \hline
\end{tabular}
\vspace{-0.5cm}
\end{table}
In the next section, we show our DSE approach to explore and generate optimized encoder implementations based on the coding modes shown in Table \ref{tab:modes_HM}. 

\section{Actor-Based Encoder}
\label{sec:model}

We optimize the mode decision process by searching for a) the best processing order and b) the best skip conditions. 
For this purpose, we are using the actor-based encoder model introduced in~\cite{Rosales16} and implemented in the MAESTRO~\cite{Rosales14} framework.
An actor-based model enables us to encapsulate the functionality of an encoding mode and its condition for evaluation in a single actor.
Also, it allow us to explore parallel (multi-core) implementations of the standard.
In this work, each of the mode evaluation functions introduced in Table \ref{tab:modes_HM} is implemented as an actor. 

As the functionality of each actor can be executed independently from other actors, the execution order can be chosen freely. This is the first degree of freedom (DoF) that we will explore in the DSE.



To be able to encode standard-compliant bit streams, not all of the modes have to be executed.
Hence, mode guards can be defined that can skip the execution of a certain actor under certain conditions that can be chosen freely, except that it has to be assured that at least one of the actors is always called.
These mode guards enabling conditional execution of modes is the second DoF that we explore automatically during DSE. 


\subsection{Best-Mode Conditional Evaluation}
\label{sec:modeGuards}

The coding mode evaluation condition can, e.g., be based on the best mode so far, on the current costs, but also on more complicated conditions depending on current block sizes or specific properties of the best mode or even the second best mode.
In our work, we choose the best mode so far as a criterion for each individual evaluation of a coding mode. The best mode so far is the mode that, among all modes that have already been tested, shows least rate-distortion costs. 
We only evaluate the next mode if the best mode so far is the mode indicated by the mode guard. Naturally, this is a highly constrained approach and more general constraints could be defined.
Nevertheless, in this work we choose this basic approach to prove that the concept has the potential to derive highly efficient encoding solutions. Furthermore, the search space for the DSE is much smaller such that efficient solutions can be found in an acceptable amount of time.

\subsection{Design Space Exploration}
\label{sec:DSE}
As an optimization algorithm for our DSE we choose the multi-objective evolutionary algorithm presented in~\cite{deb2002fast}.
In order to perform such a DSE, a so-called genetic representation of the design space must be defined, see~\cite{opt4j}.
Such a representation for the first degree of freedom can be found straightforwardly:
we define a vector $\mathcal{O}$ with $11$ entries, where the position indicates the mode testing order.
In this vector, each value from $0$ to $10$ (cf. Table \ref{tab:modes_HM}) appears once, and the order of execution corresponds to the order of appearance.

\begin{figure}
\centering
\includegraphics[width=0.48\textwidth]{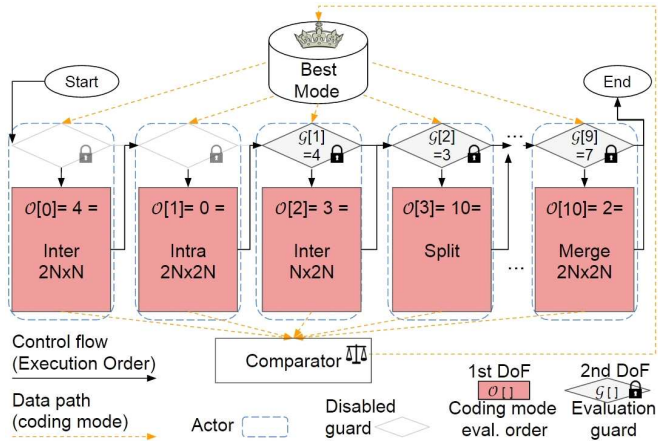}
\vspace{-.2cm}
\caption{Mode order and mode guard exploration. The coding mode order (first DoF) and the early skip conditions (second DoF) are encoded as vectors $\mathcal{O}(d)$ and $\mathcal{G}(d)$, respectively, where one vector is defined for each depth $d$.
The evaluation of a coding mode is performed if the best mode so far is equal to the mode indexed by $\mathcal{G}$.
}
\label{fig:actorModel}
\vspace{-.5cm}
\end{figure}

For the second DoF, we define another vector $\mathcal{G}$ that contains elements $g\in \{0,...,10\}$, where the element $g$ refers to the best mode so far as defined in Table \ref{tab:modes_HM}.
In our proposal, we decide to only execute the actor if the best mode so far is the mode which is indicated by the corresponding position in vector $\mathcal{G}$. E.g., referring to Fig. \ref{fig:actorModel}, $\mathcal{G}[1]=4$ means that InterNx2N is only tested if the best mode so far is mode $4$ (Inter2NxN). 
Hence, the vector only needs to have nine entries because when starting the search, at least two different modes have to be tested to be able to determine a currently best mode. 

Note that for both vectors $\mathcal{O}$ and $\mathcal{G}$, we allow different solutions for each depth $d$. That means that for a smaller block size, a different order and different guards can be used than for a bigger block size. This can be helpful as, e.g., low QPs tend to favor smaller block sizes such that bigger sizes can be skipped more often. 

Figure \ref{fig:actorModel} shows the execution flow of a solution found by the DSE. The order given by vector $\mathcal{O}$ is represented by the red boxes. On top of the boxes, the mode guards (vector $\mathcal{G}$) are represented by diamonds. After testing a coding mode, the comparator determines the currently best mode (top of the figure) which is then used to evaluate the guards. 

We choose an iterative optimization approach where all explored non-dominated solutions are stored in an archive.
To determine the performance of a solution, we encode six different sequences 
at constant quantization parameters (QPs) 10, 20, 30 and 40, and determine the values of the four objectives PSNR, rate, encoding time, and decoding energy.
We choose different QPs as they may lead to different optimal solutions.
The results presented are obtained after 200 iterations for each QP. 
The determination of the objectives is introduced in the next section. 

\section{Objectives}
\label{sec:obj}
As optimization objectives for the DSE we choose the traditionally optimized criteria distortion and rate, as a third criterion the encoding time, and finally the decoding energy. The calculation of rate and quality (PSNR) is readily implemented in the HM-encoder solution and can be directly used for DSE.
To measure encoding time, we instrument the code with a high-precision \texttt{clock()}-function taken from the C++ standard library.
The actor based HM-encoder used for the DSE allows to evaluate the parallel execution of the encoding process providing a lot of flexibility in the design space.
However, this means that the encoder is a major refactorization of the original HM code (which means that the code is changed significantly without changing functionality) such that encoding times measured during DSE are approximations.
Hence, we generated an ``unwrapped'' encoder for each selected solution that, in comparison to the baseline HM encoder, only differs in terms of mode evaluation order and mode guards.
The order and the guards (implemented as \texttt{if}-clauses) are generated from the vectors $\mathcal{O}$ and $\mathcal{G}$ as defined in Section \ref{sec:DSE}. 



The decoding energy is computed as follows: during DSE, we estimate the decoding energy by analyzing the coded bit stream for so-called bit stream features $f$ that are used to estimate the complete decoding energy $\hat E$ as 
\begin{equation}
\hat E = \sum_f n_f \cdot e_f, 
\end{equation}
where $n_f$ is the number of occurences of a certain feature and $e_f$ the feature's specific energy. A feature is, e.g., the intraframe prediction process of a block of a certain size that can a) be counted and b) requires a rather constant amount of processing energy during decoding. We use the energy estimator that was presented in \cite{Herglotz16b} and that is publicly available at \cite{denesto}. This method makes sure that the DSE does not depend on real measurements. 

For evaluation, we performed real measurements of the decoding process to prevent estimation errors. The decoding energy was measured for FFmpeg software decoding \cite{FFmpeg} on a Pandaboard \cite{Panda} which has a smartphone-like architecture using an ARM processor. The measurement setup is the same as was presented in detail in \cite{Herglotz16b}. 

\section{Evaluation}
\label{sec:eval}
In this section, we show that the proposed approach is able to find not only one, but sets of efficient encoder solutions under multiple objectives. Therefore, in the first subsection, we discuss results from the training process (the DSE). In the second subsection, we validate five different solutions taken from the DSE in a realistic scenario.  By performing training and validation on different sets of input sequences as shown in Table \ref{tab:sequences}, we show that results obtained by the DSE are valid in general. 
Therefore, we take a different video sequence from each class of the JCT-VC common test conditions \cite{Bossen13} and evaluate encoding time, decoding energy, bitrate, and quality in terms of Y-PSNR for the original HM and the solutions selected from the DSE. As an encoder configuration, lowdelay was used. Finally, the third subsection explicitly presents two showcase solutions and interprets their algorithmic behavior. 
\begin{table}[t]
\centering
\caption{Sequences (classes A-F) from the common test conditions \cite{Bossen13}. }
\label{tab:sequences}
\vspace{-.35cm}
\small
\begin{tabular}{p{.4cm}| p{4cm}| p{3.3cm}}
\hline
Cl. & DSE Training Video Sequences & Test Video Sequences \\
&[Frames]&[Frames]
\\
 \hline
 A& PeopleOnStreet 2560x1600[4] & Traffic 2560x1600[4]\\
 B& BQTerrace 1920x1080[4]& ParkScene 1920x1080[4]\\
 C& BQMall 832x480[10] & PartyScene 832x480[10]\\
 D& BasketballPass 416x240[10] & BlowingBubbles 416x240[10]\\
 E& FourPeople 1280x720[4] & Johnny 1280x720[4]\\
 F& BasketballDrillText 832x480[10] & SlideEditing 1280x720[10]\\
 \hline
\end{tabular}
\end{table}

\subsection{Training Results}

\begin{figure}[t]
\centering
\psfrag{000}[c][r]{$3\%$}
\psfrag{001}[c][c]{}
\psfrag{002}[c][r]{$0\%$}
\psfrag{003}[c][c]{$0$}
\psfrag{004}[c][c]{$10$}
\psfrag{005}[c][c]{$20$}
\psfrag{006}[c][c]{$30$}
\psfrag{007}[c][c]{$40$}
\psfrag{008}[c][c]{$50$}
\psfrag{009}[r][r]{$80$}
\psfrag{010}[r][r]{$60$}
\psfrag{011}[r][r]{$40$}
\psfrag{012}[r][r]{$20$}
\psfrag{013}[r][r]{$0$}
\psfrag{014}[b][t]{Enc. time savings (\%)}
\psfrag{015}[t][b]{Rate increase (\%)}
\psfrag{016}[c][c]{\footnotesize{P3}}
\psfrag{017}[c][c]{\footnotesize{P2}}
\psfrag{018}[c][c]{\footnotesize{PB}}
\psfrag{019}[c][c]{\footnotesize{PE}}
\psfrag{020}[c][c]{\footnotesize{P1}}
\psfrag{021}[l][l]{Energy}
\psfrag{022}[l][l]{savings}
\includegraphics[width=0.5\textwidth]{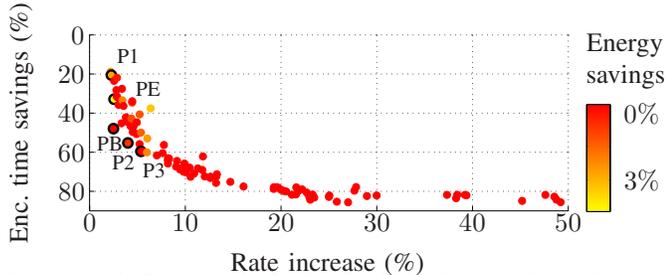}
\vspace{-0.6cm}
\caption{Results from the DSE for QP 20. The x-axis shows the rate increase, the y-axis the encoding time savings normalized to standard HM encoding. The color represents the decoding energy savings. Each point represents one Pareto-optimal solution found by the DSE, where the values are averaged over all training sequences. 
The Pareto-front of the bitrate-time trade-off is located on the bottom left outer hull of the points. Solutions chosen for validation are highlighted by black edged markers. }
\label{fig:results_QP40}
\vspace{-0.5cm}
\end{figure}

The results of the training process for QP 20 are depicted in Figure \ref{fig:results_QP40}. Results from the DSEs for the other QPs are not shown as the distribution of the points is similar. 
In this plot, rate increases, encoding time savings, and energy savings are displayed relative to the standard HM encoder.
Each marker corresponds to a Pareto-optimal (non-dominated) solution found during the DSE, where $100$ solutions are displayed.
The PSNR (which is approximately constant for the fixed QP) is not depicted to enhance visibility. We can see that a shorter encoding time generally causes an increase in bitrate. Furthermore, we can see that rate is higher than HM for all solutions. This is caused by the implementation of the guards where many of the possible modes are not evaluated. Highest decoding energy savings are close to $3\%$, as indicated by the color of the markers. 


To show more properties of some non-dominated solutions, we manually choose five explicit solutions for further analysis. In Figure \ref{fig:results_QP40}, they are marked by black edged markers. They are chosen based on the following considerations: the first selected solution aims at minimizing the bitrate (leftmost marker PB), 
the second selected solution is one minimizing the decoding energy (yellow marker PE), 
the last three solutions denote Pareto-optimal points for different trade-offs between encoding time, decoding energy, bitrate, and quality (the remaining markers P1, P2, P3). We discuss these points in the next subsection in detail.

\subsection{Validation}

As our experiments indicate that the best mode orders and guards differ strongly depending on the QP, we combine solutions from the four independent DSEs for QPs 10, 20, 30, and 40 for the four considered points PB, PE, P1, P2, and P3.
A combined solution corresponds to a QP-dependent coding order and mode guard realization, which, in a real application, could be chosen depending on the user defined input QP. 
The selection of the combinations was done by choosing points with a similar rate-distortion-time-energy performance. 

\begin{figure}[t]
\centering
\psfrag{000}[r][l]{$0$}
\psfrag{001}[r][l]{$2$}
\psfrag{002}[r][l]{$4$}
\psfrag{003}[r][l]{$6$}
\psfrag{004}[r][l]{$8$}
\psfrag{005}[r][l]{}
\psfrag{006}[r][c]{}
\psfrag{007}[r][c]{$60$}
\psfrag{008}[r][c]{$40$}
\psfrag{009}[r][c]{$20$}
\psfrag{010}[r][c]{$0$}
\psfrag{011}[r][c]{}
\psfrag{012}[b][t]{Enc. time savings ($\%$)}
\psfrag{013}[t][c]{BD-rate ($\%$)}
\psfrag{021}[c][c]{\tiny{PB}}
\psfrag{014}[l][c]{\small{S17}}
\psfrag{015}[l][c]{\small{S14}}
\psfrag{016}[l][c]{\small{S27}}
\psfrag{017}[c][c]{\small{P1}}
\psfrag{018}[l][c]{\small{P2}}
\psfrag{019}[r][b]{\small{P3}}
\psfrag{020}[c][c]{\small{PE}}
\psfrag{021}[c][c]{\small{PB}}
\includegraphics[width=0.5\textwidth]{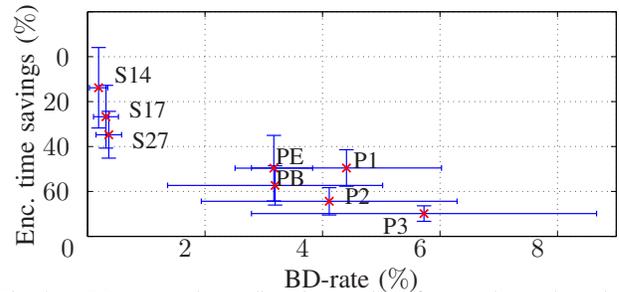}
\vspace{-0.6cm}
\caption{BD-rates and encoding time savings for tested encoder solutions. The reference points with solutions from literature are shown on the top left, the remaining points correspond to our suggestions. }
\label{fig:BD_rate}
\vspace{-0.4cm}
\end{figure}

\begin{figure}[t]
\centering
\psfrag{000}[r][l]{$-6$}
\psfrag{001}[r][l]{$-5$}
\psfrag{002}[r][l]{$-4$}
\psfrag{003}[r][l]{$-3$}
\psfrag{004}[r][l]{$-2$}
\psfrag{005}[r][l]{$-1$}
\psfrag{006}[r][l]{$0$}
\psfrag{007}[r][l]{$1$}
\psfrag{008}[r][l]{$2$}
\psfrag{009}[r][l]{$3$}
\psfrag{010}[r][c]{}
\psfrag{011}[r][c]{$60$}
\psfrag{012}[r][c]{$40$}
\psfrag{013}[r][c]{$20$}
\psfrag{014}[r][c]{$0$}
\psfrag{015}[r][c]{}
\psfrag{016}[b][t]{Enc. time savings ($\%$)}
\psfrag{017}[t][c]{BD-energy ($\%$)}
\psfrag{018}[c][c]{\tiny{S17}}
\psfrag{019}[c][c]{\tiny{S14}}
\psfrag{020}[c][c]{\tiny{S27}}
\psfrag{021}[c][c]{\tiny{P1}}
\psfrag{022}[c][c]{\tiny{P2}}
\psfrag{023}[c][c]{\tiny{P3}}
\psfrag{024}[c][c]{\tiny{PE}}
\psfrag{025}[c][c]{\tiny{PB}}
\psfrag{018}[c][c]{\small{S17}}
\psfrag{019}[r][c]{\small{S14}}
\psfrag{020}[c][c]{\small{S27}}
\psfrag{021}[c][c]{\small{P1}}
\psfrag{022}[c][b]{\small{P2}}
\psfrag{023}[c][c]{\small{P3}}
\psfrag{024}[c][c]{\small{PE}}
\psfrag{025}[c][c]{\small{PB}}
\includegraphics[width=0.5\textwidth]{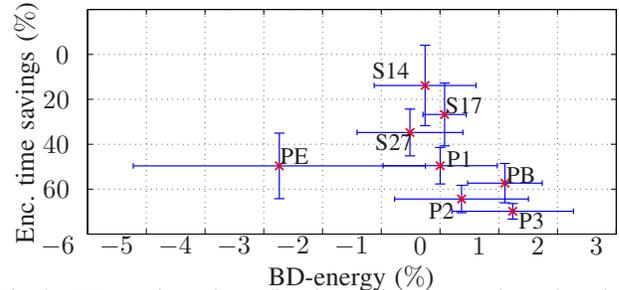}
\vspace{-0.6cm}
\caption{BD-energies and encoding time savings for tested encoder solutions.  }
\label{fig:BD_energy}
\vspace{-0.5cm}
\end{figure}

To compare the performance of these combined solutions, we calculate the Bj{\o}ntegaard delta bitrates (BD-rates) \cite{Bjonte01}, mean encoding times, and Bj{\o}ntegaard-Delta decoding energies (BD-energy) \cite{Herglotz16a} for the five considered points and compare them to three manually optimized encoding solutions from literature \cite{Vanne14} (points S14, S17 and S27). As a reference, we again take the values from the standard HM-encoder. The results are depicted in Figures \ref{fig:BD_rate} and \ref{fig:BD_energy}. 

Figure \ref{fig:BD_rate} plots the mean encoding time savings and the mean BD-rates for all considered points, where the mean of all validation sequences was taken. The horizontal and the vertical lines indicate the standard deviations. We can see that the reference solutions have a very low loss in bitrate and that they save almost $40\%$ of encoding time. The points we suggest lose more bitrate but save more than $50\%$ of decoding time. Here, especially points PB, P2, and P3 show interesting results, where the solution PB will be discussed in detail later. 

Figure \ref{fig:BD_energy} shows the encoding time savings over the BD-energy. We can see that by choosing point PE, it is possible to save $3\%$ of decoding energy. The other points show a rather low impact on the decoding energy consumption.

\begin{table*}[t]
\renewcommand{\arraystretch}{1.3}
\centering
\caption{Depth-dependent mode orders $\mathcal{O}(d)$ and guards $\mathcal{G}(d)$ for points PE and PB (QP 40). The numbers correspond to the mode indices shown in Table \ref{tab:modes_HM}. A hyphen in the guards vector indicates that the corresponding mode is always tested. }
\label{tab:PB_PE}
\vspace{-.3cm}
\tiny{
\begin{tabular}{l | l | c | c | c}
\hline
PB & 
\begin{tabular}{l}
$\mathcal{O}(0) = \{ 10,2,0,6,3,4,7,5,1,8\}$ \\ 
$\mathcal{G}(0) = \{ -,-,10,10,0,6,4,7,2,0\}$ \\
\end{tabular}
&
\begin{tabular}{l}
$\mathcal{O}(1) = \{2,3,10,0,5,4,8,7,1,6 \}$\\
$\mathcal{G}(1) = \{ -,-,3,2,3,3,4,2,0,0\}$\\
\end{tabular}
& 
\begin{tabular}{l}
$\mathcal{O}(2) = \{ 4,2,7,5,10,8,6,1,0,3\}$\\
$\mathcal{G}(2) = \{ -,-,4,2,2,5,5,4,8,10\}$\\
\end{tabular}
&
\begin{tabular}{l}
$\mathcal{O}(3) = \{ 1,3,2,4,0,9\}$\\
$\mathcal{G}(3) = \{ -,-,1,3,2,4\}$\\
\end{tabular}
 \\

\hline

PE & 
\begin{tabular}{l}
$\mathcal{O}(0) = \{ 10,1,0,5,7,6,8,3,4,2\}$ \\ 
$\mathcal{G}(0) = \{-,-,1,0,5,0,6,0,6,4 \}$ \\
\end{tabular}
&
\begin{tabular}{l}
$\mathcal{O}(1) = \{ 10,1,2,8,6,4,0,3,5,7\}$ \\ 
$\mathcal{G}(1) = \{ -,-,10,1,8,2,1,1,8,10\}$ \\
\end{tabular}
&
\begin{tabular}{l}
$\mathcal{O}(2) = \{ 1,10,0,4,3,6,2,5,7,8\}$ \\ 
$\mathcal{G}(2) = \{ -,-,1,10,10,0,3,6,2,6\}$ \\
\end{tabular}
&
\begin{tabular}{l}
$\mathcal{O}(3) = \{3,9,1,4,2,0\}$ \\ 
$\mathcal{G}(3) = \{-,-,3,1,4,1 \}$ \\
\end{tabular}
\\
 
\hline
\end{tabular}}
\vspace{-0.5cm}
\end{table*}

\subsection{Interpretation}
Finally, we would like to discuss the solutions PE and PB in more detail. The concrete realizations are shown in Table \ref{tab:PB_PE}. 
We can see that the first two modes (split $=10$ and Merge2Nx2N $=2$) are always tested on depth $d=0$. If a mode is located at the end of the order vector $\mathcal{O}$, it is less likely that it will be tested, because a single guard mode must be the best so far. 
Watching the mode order on the other depths, we find that modes $2$ and $10$ are always tested early. In this case, we can say that solution PB works well because mode 2 (Merge2Nx2N) is often chosen, which has the following two advantages: 
\begin{itemize}
\item Merge mode testing is faster than the other modes (no motion or intra mode search required), 
\item In terms of BD-rate, the merge and skip modes are highly effective for high QPs (precision of motion vectors and residual coefficients is unimportant). 
\end{itemize}

For point PE, we found that mode 1 (Inter2Nx2N) is chosen in most cases which is again caused by the mode testing order. Analyzing the resulting bit streams we found that for PE, the number of bipredicted blocks is significantly lower than for PB, which causes the energy savings. Apparently, for Merge2Nx2N, biprediction is more likely to be chosen due to the automatically generated motion vector candidates, cf. \cite{Sullivan12}, where Inter2Nx2N often favors uniprediction due to a more accurate motion search. 


Furthermore, we can observe that the variance of the proposed solutions is much higher than for the reference points. This can be explained by the fact that in the reference, only the PU splitting modes for inter coding are skipped. In contrast, our approach allows the complete skipping of basic prediction modes like intra, skip, or splitting which can cause higher deviations depending on the input sequence. 

\section{Conclusions}
\label{sec:concl}

We presented a multi-objective DSE approach that is able to find encoding solutions that exploit the trade-off between encoding speedup, decoding energy, and BD-rate.
Our automatic approach together with a simple early skip condition heuristic could determine solutions achieving speedups of more than $60\%$ and decoding energy savings of more than $3\%$. 

In future work, thanks to the actor-based model, a corresponding study could be performed for multi-core processors where the optimal encoding order could change due to potential parallel execution of coding modes.
Furthermore, the encoding process could be further split into additional actors to be able to explore early decision possibilities at an even finer granular level.

\section*{ACKNOWLEDGEMENT}
This work was financially supported by the Research Training Group 1773 ``Heterogeneous Image Systems'', funded by the German Research Foundation (DFG).

\bibliographystyle{IEEEtran}
\bibliography{IEEEabrv,literature}

\begin{thebibliography}{10}
\providecommand{\url}[1]{#1}
\csname url@samestyle\endcsname
\providecommand{\newblock}{\relax}
\providecommand{\bibinfo}[2]{#2}
\providecommand{\BIBentrySTDinterwordspacing}{\spaceskip=0pt\relax}
\providecommand{\BIBentryALTinterwordstretchfactor}{4}
\providecommand{\BIBentryALTinterwordspacing}{\spaceskip=\fontdimen2\font plus
\BIBentryALTinterwordstretchfactor\fontdimen3\font minus
  \fontdimen4\font\relax}
\providecommand{\BIBforeignlanguage}[2]{{%
\expandafter\ifx\csname l@#1\endcsname\relax
\typeout{** WARNING: IEEEtran.bst: No hyphenation pattern has been}%
\typeout{** loaded for the language `#1'. Using the pattern for}%
\typeout{** the default language instead.}%
\else
\language=\csname l@#1\endcsname
\fi
#2}}
\providecommand{\BIBdecl}{\relax}
\BIBdecl

\bibitem{Sullivan12}
G.~Sullivan, J.~Ohm, W.-J. Han, and T.~Wiegand, ``Overview of the high
  efficiency video coding {(HEVC)} standard,'' \emph{IEEE Transactions on
  Circuits and Systems for Video Technology}, vol.~22, no.~12, pp. 1649 --1668,
  Dec. 2012.

\bibitem{HM}
\BIBentryALTinterwordspacing
{ITU\slash ISO\slash IEC}. {HEVC Test Model HM}. [Online]. Available:
  \url{https://hevc.hhi.fraunhofer.de/}
\BIBentrySTDinterwordspacing

\bibitem{jctvc_esd}
J.~Yang, J.~Kim, K.~Won, H.~Lee, and B.~Jeon, ``Early skip detection for
  {HEVC},'' \emph{Joint Collaborative Team on Video Coding (JCT-VC), Document
  JCTVC-G543, Geneva, Switzerland}, 2011.

\bibitem{jctvc_ecu}
K.~Choi, S.-H. Park, and E.~S. Jang, ``Coding tree pruning based {CU} early
  termination,'' \emph{Joint Collaborative Team on Video Coding (JCT-VC),
  Document JCTVC-F092, Torino, Italy}, 2011.

\bibitem{jctvc_cfm}
R.~H. Gweon, Y.-L. Lee, and J.~Lim, ``Early termination of {CU} encoding to
  reduce {HEVC} complexity,'' \emph{Joint Collaborative Team on Video Coding
  (JCT-VC), Document JCTVC-F045, Torino, Italy}, 2011.

\bibitem{Zhang14b}
H.~Zhang and Z.~Ma, ``Fast intra mode decision for high efficiency video coding
  ({HEVC}),'' \emph{IEEE Transactions on Circuits and Systems for Video
  Technology}, vol.~24, no.~4, pp. 660--668, April 2014.

\bibitem{Heindel16}
A.~Heindel and A.~Kaup, ``Fast exclusion of angular intra prediction modes in
  {HEVC} using reference sample variance,'' in \emph{Proc. IEEE International
  Symposium on Circuits and Systems (ISCAS)}, Montreal, Canada, May 2016, pp.
  2675--2678.

\bibitem{Cao13}
Y.~Cao and S.~Goto, ``A mode filtering algorithm for accelerating {HEVC FME},''
  in \emph{IEEE 15th International Workshop on Multimedia Signal Processing
  (MMSP)}, Sept 2013, pp. 218--223.

\bibitem{shang2015fast}
X.~Shang, G.~Wang, T.~Fan, and Y.~Li, ``Fast cu size decision and pu mode
  decision algorithm in {HEVC} intra coding,'' in \emph{Proc. IEEE
  International Conference on Image Processing (ICIP)}, Qu\'ebec City, Canada,
  2015, pp. 1593--1597.

\bibitem{Vanne14}
J.~Vanne, M.~Viitanen, and T.~Hamalainen, ``Efficient mode decision schemes for
  {HEVC} inter prediction,'' \emph{IEEE Trans. on Circuits and Systems for
  Video Technology}, vol.~24, no.~9, pp. 1579--1593, Sept 2014.

\bibitem{Herglotz16b}
C.~Herglotz, D.~Springer, M.~Reichenbach, B.~Stabernack, and A.~Kaup,
  ``Modeling the energy consumption of the {HEVC} decoding process,''
  \emph{accepted for IEEE Transactions on Circuits and Systems for Video
  Technology}, 2016.

\bibitem{Herglotz16a}
C.~Herglotz and A.~Kaup, ``Joint optimization of rate, distortion, and decoding
  energy for {HEVC} intraframe coding,'' in \emph{Proc. IEEE International
  Conference on Image Processing (ICIP)}, Phoenix, USA, 2016.

\bibitem{Rosales16}
R.~Rosales, C.~Herglotz, M.~Gla{\ss}, J.~Teich, and A.~Kaup, ``Analysis and
  exploitation of {CTU}-level parallelism in the {HEVC} mode decision process
  using actor-based modeling,'' in \emph{Proc. 29th Int. Conf. on Architecture
  of Computing Systems (ARCS)}, N\"urnberg, Germany, Apr 2016, pp. 263--276.

\bibitem{deb2002fast}
K.~Deb, A.~Pratap, S.~Agarwal, and T.~Meyarivan, ``A fast and elitist
  multiobjective genetic algorithm: {NSGA-II},'' \emph{Evolutionary
  Computation, IEEE Transactions on}, vol.~6, no.~2, pp. 182--197, 2002.

\bibitem{Rosales14}
R.~Rosales, M.~Glass, J.~Teich, B.~Wang, Y.~Xu, and R.~Hasholzner,
  ``{MAESTRO}-- {H}olistic actor-oriented modeling of nonfunctional properties
  and firmware behavior for {MPSoCs},'' \emph{ACM Trans. Des. Autom. Electron.
  Syst.}, vol.~19, no.~3, pp. 23:1--23:26, Jun. 2014.

\bibitem{opt4j}
M.~Lukasiewycz, M.~Gla{\ss}, F.~Reimann, and J.~Teich, ``{Opt4J - A Modular
  Framework for Meta-heuristic Optimization},'' in \emph{Proceedings of the
  Genetic and Evolutionary Computing Conference (GECCO 2011)}, Dublin, Ireland,
  2011, pp. 1723--1730.

\bibitem{denesto}
\BIBentryALTinterwordspacing
{Decoding Energy Estimation Tool (DENESTO)}. (2016). [Online]. Available:
  \url{http://lms.lnt.de/denesto}
\BIBentrySTDinterwordspacing

\bibitem{FFmpeg}
(2016) {FFmpeg}. http://ffmpeg.org/. Accessed 2016-10-01.

\bibitem{Panda}
(2015) pandaboard.org. http://pandaboard.org/. Accessed 2015-10-07.

\bibitem{Bossen13}
F.~Bossen, ``Common test conditions and software reference configurations,''
  document JCTVC-L1100, ITU-T VCEG and ISO/IEC MPEG (JCT-VC), Geneva,
  Switzerland, Jan. 2013.

\bibitem{Bjonte01}
G.~Bj{\o}ntegaard, ``Calculation of average psnr differences between rd
  curves,'' VCEG-M33, Austin, TX, USA, document, Apr. 2001.

\end{thebibliography}

\end{document}